# Evaluation of “As-Intended” Vehicle Dynamics using the Active Inference Framework

Kazuharu Kidera[1[0009-0008-8969-6468]], Takuma Miyaguchi[2[0009-0006-6229-4317]], and Hideyoshi Yanagisawa[2[0000-0002-0175-7537]]

[1] Honda R&D Co., Ltd., 4630 Shimotakanezawa, Haga-machi, Haga-gun, Tochigi 321-3393, Japan
[2] The University of Tokyo, 7-3-1 Hongo, Bunkyo-ku, Tokyo 113-8656, Japan
kazuharu_kidera@jp.honda

**Abstract.** We constructed a computational model of the driver’s brain for steering tasks using the active inference framework, grounded in the free energy principle—a theory from computational neuroscience. This model enables quantitative estimation of how accurately the brain learns vehicle dynamics and performs appropriate steering, using a measure called variational free energy. Through driving simulator experiments, we observed strong correlations between variational free energy and both expert drivers’ subjective “as-intended” scores and general participants’ objective control performance. These results suggest that variational free energy provides a promising quantitative metric for evaluating whether a vehicle behaves “as-intended.”

## 1 Introduction

### 1.1 Research Objective

In the development of vehicle dynamics performance, static design parameters are typically set based on accumulated experience and established engineering theories. Basic performance is then validated using physically measurable or simulation-based dynamic indicators, such as steering response [1]. However, the final adjustment of these parameters still relies heavily on subjective evaluations by expert drivers—assessments that are often neither objective nor quantitative. A fundamental question remains: what does it truly mean for a driver to operate a vehicle “as-intended”? We address this issue by applying the free energy principle [2] and active inference [3]—neuroscientific theories—as a novel approach to understanding and evaluating “as-intended” control.

### 1.2 Free Energy Principle and Active Inference

The free energy principle is a theory that explains the brain's perception, learning, and action in a unified manner. By interacting with the external environment through the body, the brain constructs a model of the world known as the generative model. The process of building this model corresponds to learning and action, while perception is realized as Bayesian inference through the model. These processes are achieved by minimizing a quantity called variational free energy (VFE), which represents the modeling error [3].

Active inference builds on this principle by providing a theoretical framework for planning actions to achieve goals while continuously updating the generative model [3]. It can be implemented in a simulation environment. By modeling driver behavior within this framework, the modeling error of vehicle dynamics in the driver's brain can be quantified as VFE (see Fig. 1, modified from [3]).

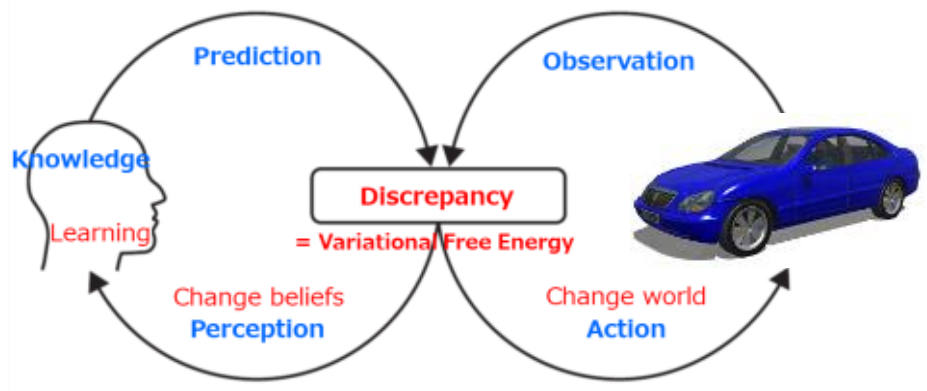

**Fig. 1.** A conceptual illustration of how perception and action minimize VFE, a measure of the discrepancy between the brain's generative model and the external world.

## 2 Methodology

### 2.1 Computational Model of the Brain in Steering Tasks

This study focuses on a steering task involving a car traveling at a constant speed of 100 km/h. Vehicle dynamics were simulated using CarSim, developed by Mechanical Simulation Corporation, now part of Applied Intuition, Inc. [4]. The driving course, 10 meters wide, includes three corners—left, right, and left—and a straight section. The objective is to follow the centerline as closely as possible. Fig. 2 shows both the layout of this course and the configuration of the driving simulator used in the experiment for this study. The CarSim model incorporates complex dynamics such as tire nonlinearity, mechanical friction, and stroke velocity-dependent damping, capturing realistic characteristics of vehicle behavior.

The first step involves constructing a self-learning driving behavior model using the active inference framework. This enables the simulation of information processing that occurs in the driver's brain. Since the vehicle dynamics are too complex to describe analytically, we adopted a discrete-time active inference model, which avoids the need for explicit motion and observation equations. In this framework, the generative model is formulated as a partially observable Markov decision process (POMDP) based on mean-field approximation, as illustrated in Fig. 3 [3, 5]. It represents how the hidden state vector—representing the firing states of neuronal populations—evolves over time as a probability distribution. The tensors and vectors used in this generative model are summarized in Table 1.

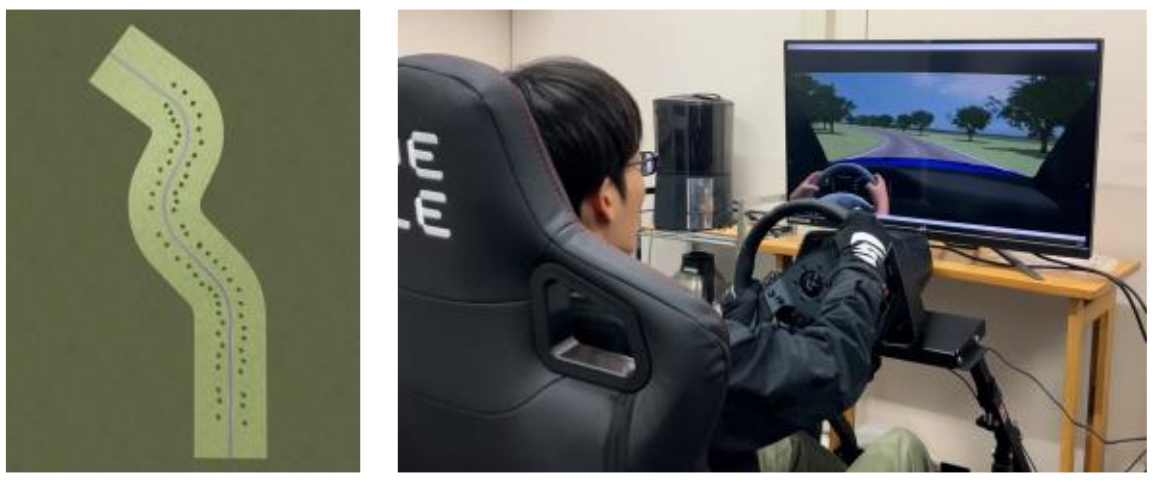

**Fig. 2.** Top-down view of the driving course and simulator setup used to evaluate "as-intended" vehicle control.

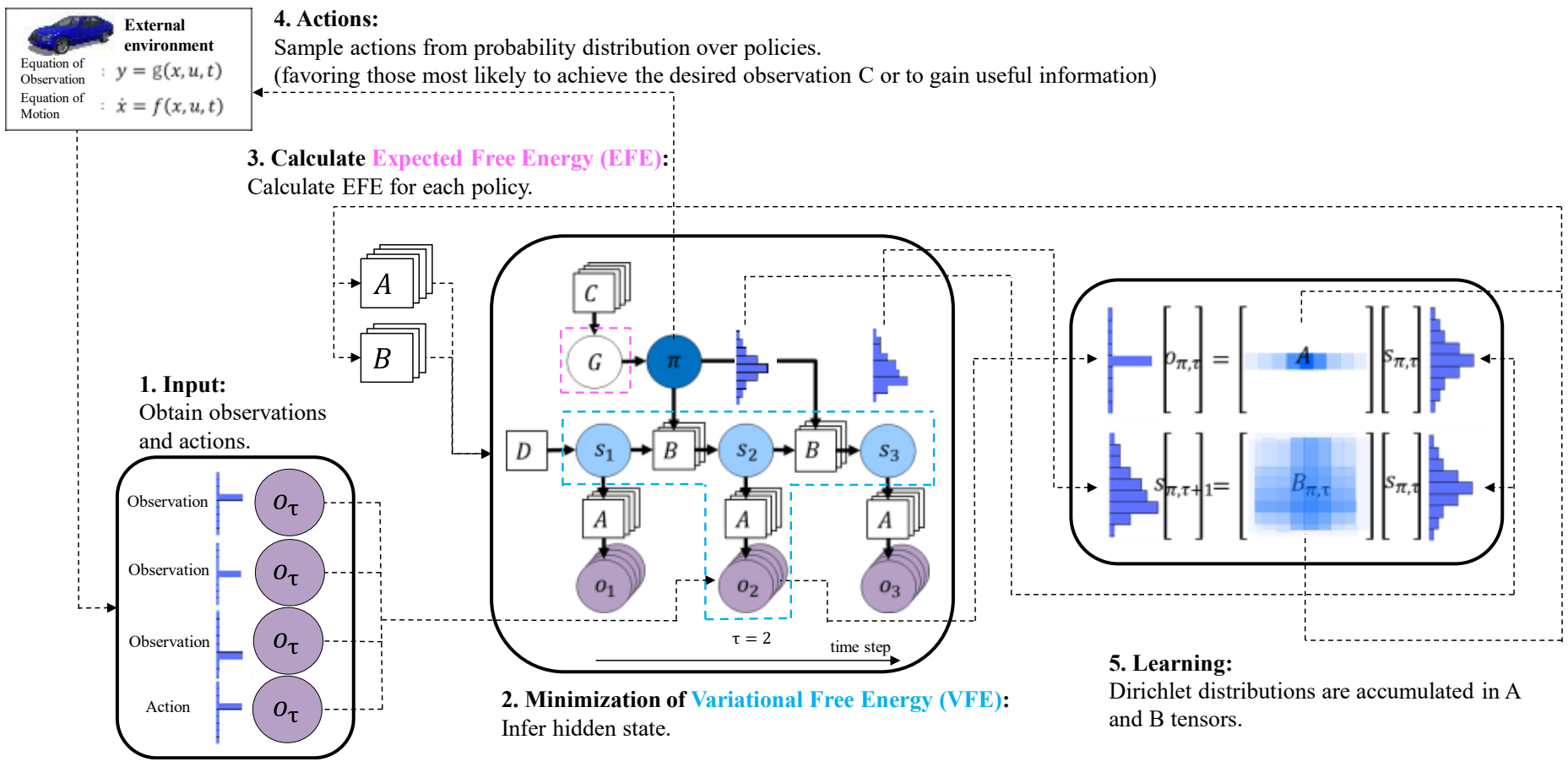

**Fig. 3.** The active inference framework consisting of: 1. Observation, 2. inference of hidden states, 3. evaluation of policies, 4. action selection, and 5. learning.

**Table 1.** Components of the generative model in the active inference framework.

| Component | Descriptions and Configurations |
|---|---|
| Observation<br>$o_\tau$ | One-hot vectors representing specific observed values, with the following dimensions for each modality:<br>- modality[0]: Lateral deviation from the centerline at 15 m ahead (−10 to +10 m, 64 bins)<br>- modality[1]: Area error with respect to the centerline up to 15 m ahead (−160 to +160 m², 64 bins)<br>- modality[2]: Steering angle (−120 to +120 deg, 49 bins)<br>- modality[3]: Steering torque feedback (−30 to +30 Nm, 64 bins)<br>- modality[4]: Vehicle yaw rate (−40 to +40 deg./s, 64 bins)<br>- modality[5]: Vehicle lateral acceleration (−1.6 to +1.6 G, 64 bins) |
| Hidden state<br>$s_\tau$ | Categorical distribution with the following dimension:<br>- factor[0]: 36 |
| Likelihood $A$ | For each modality, the mapping from hidden states to observations has the following matrix dimensions:<br>- modality[0]: 64 × 36<br>- modality[1]: 64 × 36<br>- modality[2]: 49 × 36<br>- modality[3]: 64 × 36<br>- modality[4]: 64 × 36<br>- modality[5]: 64 × 36<br>Each matrix is column-normalized and filled with uniform values. |
| State transition $B$ | The state transition tensor has the following dimension:<br>- factor[0]: 36 × 36 × 49<br>Here, 49 corresponds to the action (steering angle) dimension. Each matrix slice corresponding to an action is column-normalized and filled with uniform values. |
| Preference $C$ | Preferences are defined for the following three modalities (the others are set as uniform categorical distributions):<br>- modality[0]: Normal distribution with a mean of 0 m and a standard deviation equivalent to 0.5 m<br>- modality[1]: Normal distribution with a mean of 0 m² and a standard deviation equivalent to 8 m²<br>- modality[3]: Normal distribution with a mean of 0 Nm and a standard deviation equivalent to 15 Nm |
| Initial state $D$ | Categorical distribution with the following dimension:<br>- factor[0]: 36<br>Initialized with a sigmoid-shaped distribution to provide a better starting point for learning than a random distribution. |

In this study, we used the JAX version of pymdp [6], an open-source Python library for active inference that supports large-scale tensor operations and parallel computing via Google's JAX [7]. We integrated pymdp, which implements the computational model of the brain and runs on WSL (Windows Subsystem for Linux), with CarSim, which represents the external environment and runs on MATLAB Simulink. Communication from CarSim to pymdp over a local Ethernet connection corresponds to sensory nerve signals, while communication in the opposite direction corresponds to motor nerve signals that control the steering angle. The cycle of the POMDP for the steering task was set to 25 Hz as a representative frequency, since a single frequency had to be chosen. Although this choice was initially based on assumptions about the timescale of motor and sensory processing, its validity is examined through the experimental results in this study. Interestingly, a study by a game developer suggests that the brain processes information during gameplay at 25–33.3 Hz [8].

Inference of hidden states, corresponding to the recognition process, was performed by minimizing VFE using marginal message passing (MMP) over the past 32 time steps of observations (approximately 1.3 seconds of sensory memory). MMP is regarded as one of the most promising variational inference algorithms in active inference [9]. The VFE of a policy $\pi$ (i.e., a sequence of actions), denoted as $F_\pi$, is defined as follows [3]:

$$F_\pi = D_{KL}[Q(\tilde{s}|\pi)||P(\tilde{o},\tilde{s}|\pi)](1)$$

In discrete-time active inference, the $F_\pi$ at time step $\tau$ is computed using MMP as [3]:

$$\boldsymbol{F}_{\pi\tau} = \boldsymbol{s}_{\pi\tau} \cdot \left( ln\, \boldsymbol{s}_{\pi\tau} - ln\, \boldsymbol{A} \cdot o_\tau - \frac{1}{2}\left(ln(\boldsymbol{B}_{\pi\tau}\boldsymbol{s}_{\pi\tau-1}) + ln\left(\boldsymbol{B}^{\dagger}_{\pi\tau+1}\boldsymbol{s}_{\pi\tau+1}\right)\right)\right)(2)$$

In active inference, the expected free energy (EFE) is computed for each policy, a softmax function is applied to obtain a probability distribution over policies, and an action—steering angle in this case—is selected from this distribution [3]. EFE can be expressed in various mathematical forms, but one formulation represents it as the sum of an epistemic value term (exploration), a pragmatic value term (exploitation), and a novelty term, as shown in Equation (3) [3].

$$G(\pi) = -E_{Q(\tilde{o}|\pi)}\left[D_{KL}[Q(\tilde{s}|\pi,\tilde{o})||Q(\tilde{s}|\pi)]\right] - E_{Q(\tilde{o}|\pi)}[ln\, P(\tilde{o}|C)] - E_{\tilde{Q}(\tilde{o},\tilde{s}|\pi)}\left[D_{KL}[Q(\theta|\tilde{o},\tilde{s})||Q(\theta)]\right](3)$$

Learning in the generative model follows the update rules given in Equation (4) for the A tensor, B tensor, and D vector [3].

$$\boldsymbol{a} = a + \sum_\tau o_\tau \otimes \boldsymbol{s}_\tau$$
$$\boldsymbol{b}_{\pi\tau} = b_{\pi\tau} + \sum_\tau \boldsymbol{s}_{\pi\tau} \otimes \boldsymbol{s}_{\pi\tau-1}$$
$$\boldsymbol{d} = d + \boldsymbol{s}_1(4)$$

With the above task setting and generative model, simulation confirmed that autonomous learning of the steering task was possible. Starting from a state where both the A and B tensors were flat, actions were generated through active inference, enabling the generative model (i.e., the computational model of the brain) to learn, much like an actual human would. During learning, steering behavior was initially random, and the vehicle frequently went off course due to a lack of knowledge about driving operations or vehicle dynamics. After several dozen trials, the generative model began to partially follow the target trajectory (the centerline), and after approximately 50 to 150 trials, it was able to complete the course while staying close to it. This progression reflects a typical pattern in active inference: initial dominance of the novelty term in EFE, followed by state exploration and eventual exploitation.

## 2.2 Hypothesis

During online learning, we found that VFE increases with greater deviation from the centerline and decreases as the deviation decreases (see Fig. 4). Fig. 1 illustrates that when VFE is low, the vehicle's behavior is predictable; when it is high, it becomes unpredictable. This unpredictability reduces the accuracy of action selection—steering in this task. Consequently, deviation from the centerline is closely linked to VFE. Since the desired observations (C vector) are not used in computing VFE, this relationship is a non-trivial finding demonstrated through simulation.

Hypothesis: Lower VFE reflects a higher degree of "as-intended" control in the steering task.

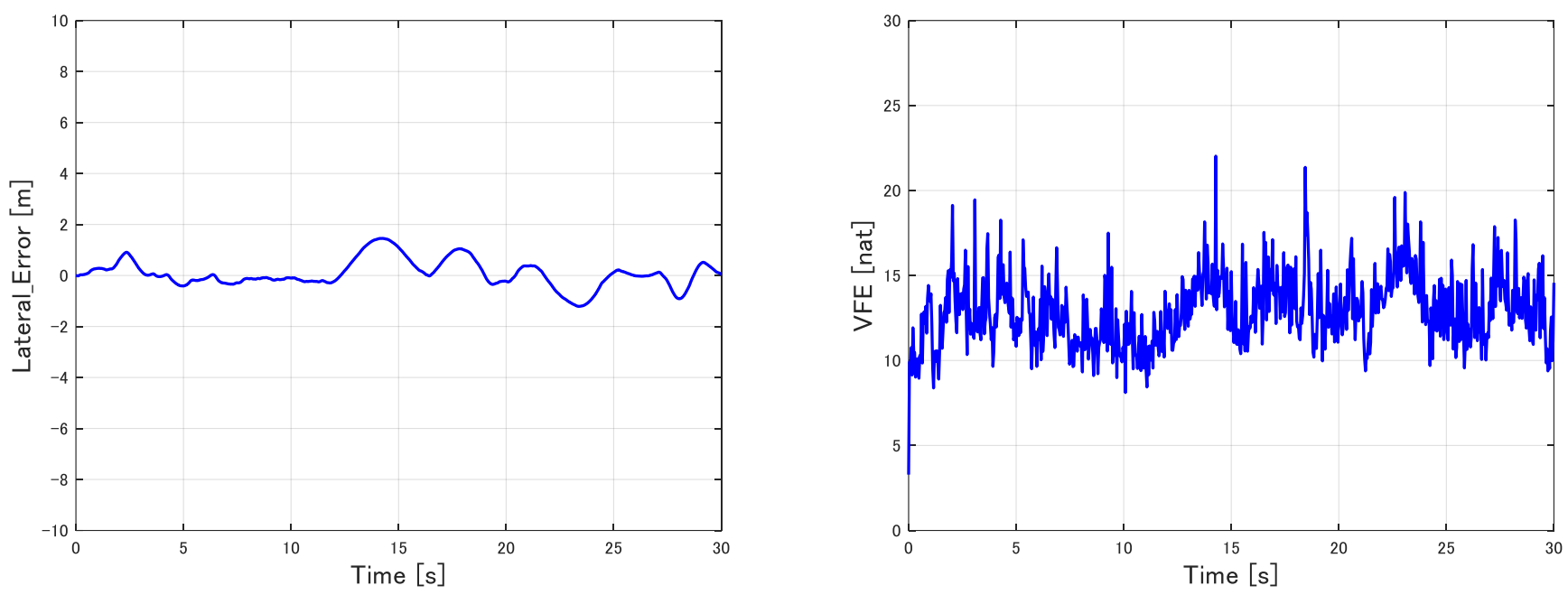

**Fig. 4.** Time-series plot of lateral deviation from the centerline and VFE during online learning of the steering task. Higher VFE is associated with greater deviation.

### 2.3 Verification of Hypothesis through Offline Learning

In this paper, we refer to autonomous learning using the active inference framework, as described in Section 2.1, as online learning. Offline learning is also possible using observation and action data from a computer-controlled or human driver, without using policy-based actions from EFE. As shown in Fig. 3 and Equation (4), the A, B, and D parameters of the generative model can be learned by sequentially inputting observation data, with the appropriate B matrix selected based on the actions.

We validated this offline approach using the course in Fig. 2 (excluding the initial 9-second straight section, added as a lead-in segment for the simulator), 12 CarSim vehicle models, and the Simulink driver model, which controls steering based on observation modalities 0 and 1 (Table 1). Fig. 5 shows the time-averaged VFE in the final trial after 15 learning iterations per vehicle. While the driver model follows the centerline reasonably well across all vehicles, the average VFE shows clear and systematic differences, reflecting each vehicle's specifications and dynamics. This supports the hypothesis in Section 2.2.

Fig. 6 shows time-series data for the vehicles with the lowest and highest average VFE. The course includes straight sections and corners with different radii. VFE rises in transitional regions where the vehicle transitions between straight and curved segments, indicating non-steady-state motion. In the low-VFE vehicle, corrective steering is minimal and observations transition smoothly. In contrast, the high-VFE vehicle shows significant corrective steering and large fluctuations during the same transitions.

| Vehicle model | | A-Class Hatchback | B-Class Hatchback | B-Class SportsCar | C-Class Hatchback | D-Class Sedan | D-Class Minivan | D-Class SUV | E-Class Sedan | E-Class SUV | Exotic SportsCar | F-Class Sedan | Full Size SUV |
|---|---|---|---|---|---|---|---|---|---|---|---|---|---|
| Drive System | | FF | FF | FR | FF | FF | FF | 4WD | 4WD | 4WD | 4WD | 4WD | 4WD |
| Weight [kg] | | 750 | 1110 | 1020 | 1270 | 1370 | 1800 | 1430 | 1650 | 1590 | 1360 | 1820 | 2257 |
| Yaw Inertia [kg m^2] | | 750 | 1343.1 | 1020 | 1536.7 | 2315.3 | 3528 | 2059.2 | 3234 | 2687.1 | 1065.2 | 4095 | 3524.9 |
| Wheelbase [mm] | | 2350 | 2600 | 2330 | 2910 | 2866 | 3000 | 2660 | 3050 | 2950 | 2650 | 3160 | 3140 |
| Average of track widths [mm] | | 1395 | 1482.5 | 1482.5 | 1675 | 1550 | 1640 | 1565 | 1600 | 1575 | 1625 | 1605 | 1737.5 |
| Center of Gravity Height [mm] | | 540 | 540 | 375 | 540 | 520 | 700 | 650 | 530 | 720 | 375 | 590 | 781 |
| Steering ratio | | 22.9 | 17.8 | 17.9 | 19.9 | 17.6 | 21.04 | 17 | 16.2 | 19.7 | 15.6 | 19.2 | 18 |
| Steering–Yaw Rate Response @ 0.2Hz | Gain [dB] | -6.16 | -7.98 | -6.88 | -10.1 | -8.26 | -13.6 | -12.0 | -11.5 | -12.4 | -3.65 | -12.8 | -11.9 |
| | Phase [deg.] | -11.1 | -5.22 | -3.01 | -3.15 | -6.82 | -2.96 | 0.015 | -1.31 | -2.25 | 0.121 | -1.82 | -3.03 |
| Steering–Lateral Accel. Response @ 0.2Hz | Gain [dB] | -32.6 | -34.3 | -33.1 | -36.3 | -34.9 | -39.9 | -38.6 | -37.8 | -39.1 | N/A | -39.1 | -38.3 |
| | Phase [deg.] | -24.0 | -14.0 | -9.87 | -9.18 | -15.4 | -11.4 | -12.1 | -9.25 | -12.7 | N/A | -7.03 | -14.1 |
| Car Sim precomputed average VFE [nat] | | 3.13 | 2.95 | 2.98 | 2.82 | 3.30 | 3.96 | 4.24 | 3.02 | 3.40 | 2.50 | 2.62 | 3.81 |

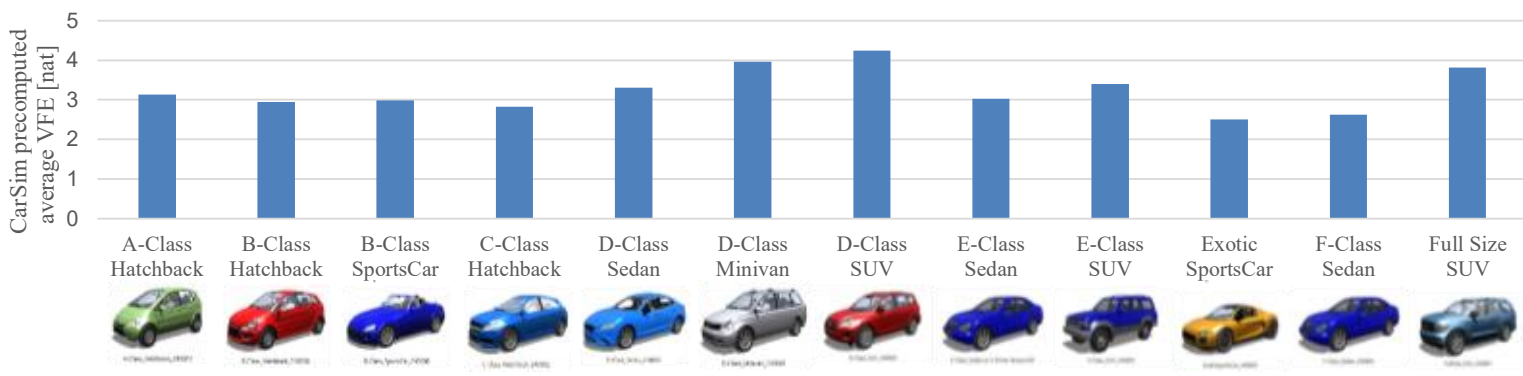

**Fig. 5.** Time-averaged VFE obtained through offline learning using the Simulink driver model for 12 CarSim vehicle models. Vehicle parameters and dynamic characteristics generally considered favorable to handling are shown in deeper blue; unfavorable ones are shown in deeper red.

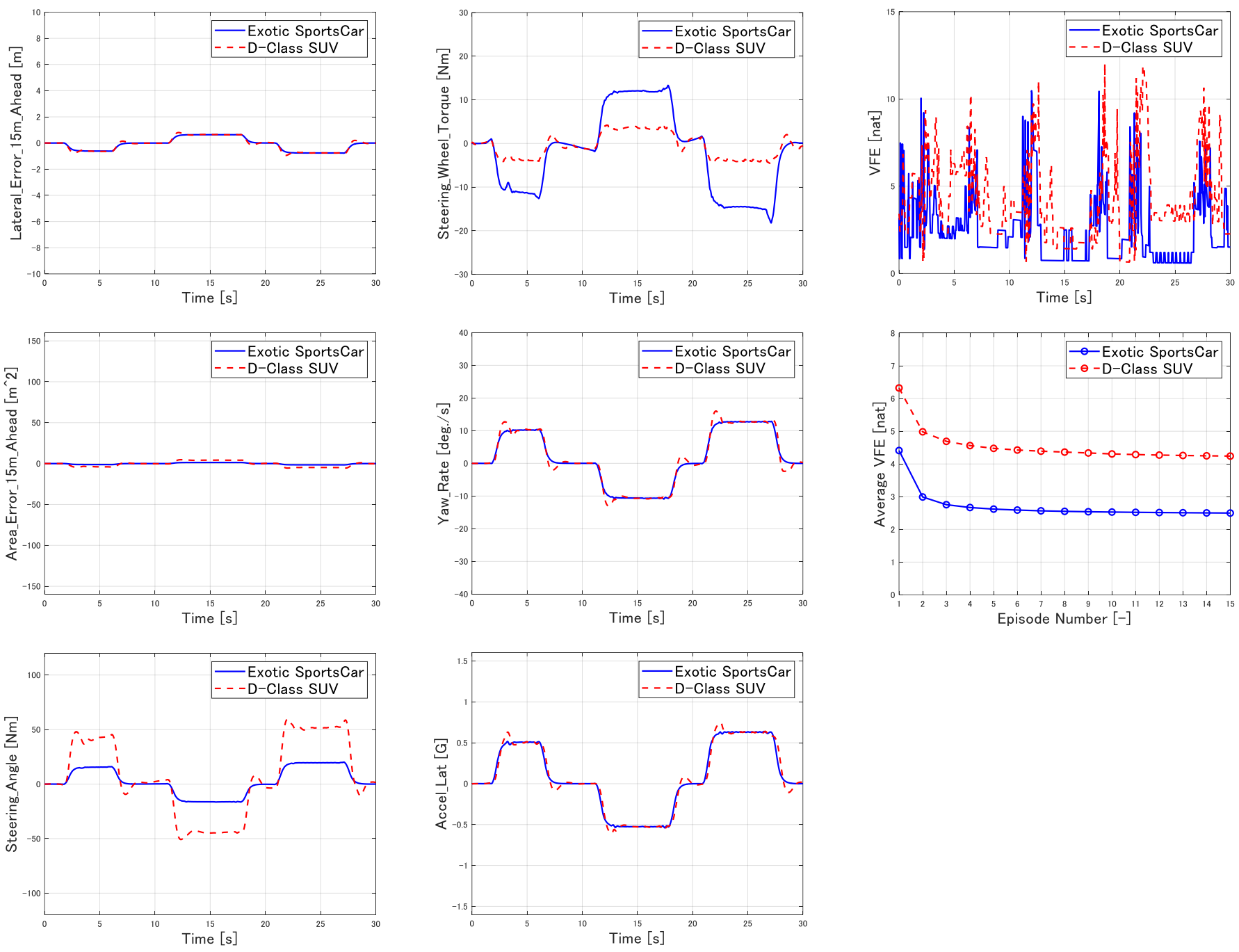

**Fig. 6.** Time-series plots of observations for each modality in Table 1. The right side displays the VFE over time in the final trial and the progression of time-averaged VFE over 15 trials.

In the active inference framework, this can be interpreted as follows. The brain learns both the frequency of co-occurrence between observations and hidden states at each time step, and the transitions between hidden states over time, as described in Equation (4). When these relationships become complex—as in vehicles with high VFE—the modeling error of the generative model (i.e., VFE) tends to increase. In such cases, achieving “as-intended” steering is expected to be extremely difficult for humans. The following sections experimentally verify this hypothesis.

## 3 Experimental Verification

### 3.1 Experimental Conditions

To verify the hypothesis presented in Section 2.2, we conducted an experiment. It was approved by the Life Ethics Committee of Honda R&D Co., Ltd. (Committee number: 100HM-041H) and the Ethics Review Committee of the Graduate School of Engineering, The University of Tokyo (Approval number: KE24-76).

Participants drove the course using CarSim on a simple simulator (see Fig. 2), with the 12 vehicle models listed in Fig. 5. They were instructed to steer as close to the centerline as possible. Vehicle speed was fixed at 100 km/h in CarSim, so throttle and brake were not used. The course took 39 seconds to complete.

Each vehicle was driven five times on the course. The first two runs were excluded due to unstable driving during familiarization and difficulty in obtaining stable offline VFE. The remaining three trials were used for analysis. To ensure data reliability, 15 off-course trials were also excluded from the 300 total, in the case of the 60 general participants described later.

All trial data from each participant’s successive runs were used for continuous offline learning, ensuring that the calculated VFE reflected their full experience. The A tensor, B tensor, and D vector listed in Table 1 were initialized only once per participant at the beginning, and continuous learning then followed Equation (4).

Before the experiment, all participants were given instructions originally provided in Japanese and translated into English for this paper. They were told that the purpose of the experiment was to evaluate whether the car behaves “as-intended” when the driver has a specific intention. Driving along the centerline was used to provide a common intention among participants. However, precision was not defined, as participants had varying levels of driving skill, making it unrealistic to demand uniform accuracy.

After each run, participants performed sensory evaluations by rating how "as-intended" they felt the car behaved. Participants were asked the following questions:

- Please score how "as-intended" you felt the car was during the previous trial. The degree of "as-intended" refers to the extent to which the car behaved as you intended.
- 0: Not at all "as-intended", 10: Completely "as-intended" (11-point Likert scale)

The evaluation focused on the perceived behavior of the car, not the participants’ driving skills. Participants included:

1. One professional expert driver with extensive experience in vehicle development.
2. Sixty general participants holding Japanese driver's licenses.

The expert driver's evaluations were considered reliable, whereas those by general participants were potentially less so. Accordingly, the expert driver evaluated all 12 vehicles in a random order specified by the experimenter. Each general participant drove one vehicle randomly assigned by the experimenter, and only the evaluation for that vehicle was used for analysis. Five participants were assigned to each vehicle model, ensuring balanced driving experience across models.

### 3.2 Results of the Experiment by Expert Driver

Fig. 7 (left) shows the relationship between the expert driver's "as-intended" scores and the average VFE obtained through offline learning using the CarSim Simulink driver's data (hereafter, CarSim precomputed average VFE; see Fig. 5). A strong negative correlation was found ($r$=-0.80, $p$<0.001), strongly supporting the hypothesis in Section 2.2. Fig. 7 (right) shows the same relationship using the expert driver's own driving data, which also revealed a relatively strong negative correlation ($r$=-0.67, $p$<0.001), further reinforcing the hypothesis.

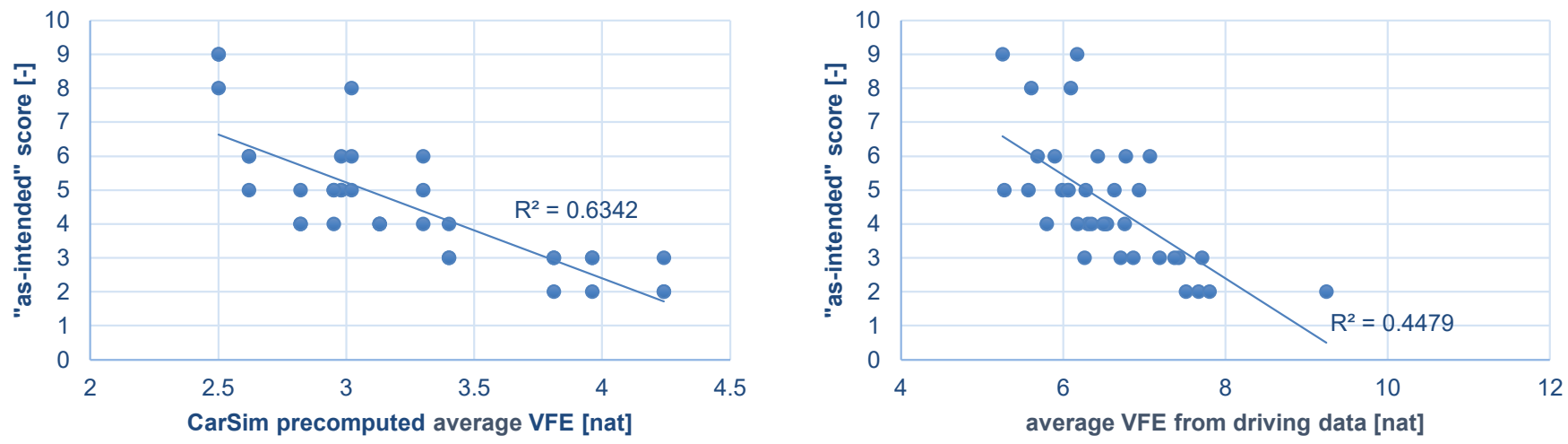

**Fig. 7.** Relationship between "as-intended" scores and average VFE values. Left: VFE from CarSim Simulink driver. Right: VFE from expert driver's own data.

The slightly weaker correlation in the expert driver's own data (r=-0.67), compared to the CarSim data (r=-0.80), may partly reflect the influence of learning order: offline learning for the expert driver was performed continuously across all vehicles, whereas the CarSim data were learned independently for each vehicle. The stronger correlation with the CarSim data suggests that the expert driver's "as-intended" scores were based on an absolute internal standard, unaffected by the driving order.

As shown in Fig. 8 (left), the expert driver exhibited almost no variation in control performance—measured as the time-averaged distance from the centerline—across different vehicle models, despite variations in the CarSim precomputed average VFE. The expert driver's steering control was comparable to that of the Simulink driver, which likely underlies the strong correlations in Fig. 7 (left) and suggests that control performance does not determine the "as-intended" score.

Fig. 8 (middle) suggests that the expert driver may have based their "as-intended" scores on the amount of corrective steering required. Vehicles with higher CarSim precomputed average VFE required greater corrective steering, as quantified by the time-averaged magnitude of steering angle components above a 1.27 Hz cutoff frequency, extracted using a washout filter. This measure showed a strong positive correlation with VFE ($r$=0.88, $p$<0.001). Furthermore, as shown in Fig. 8 (right), this corrective steering effort exhibited a strong negative correlation with the "as-intended" score ($r$ =-0.79, $p$<0.001). These results suggest that the expert driver, either consciously or unconsciously, perceived task difficulty and adjusted their steering effort accordingly.

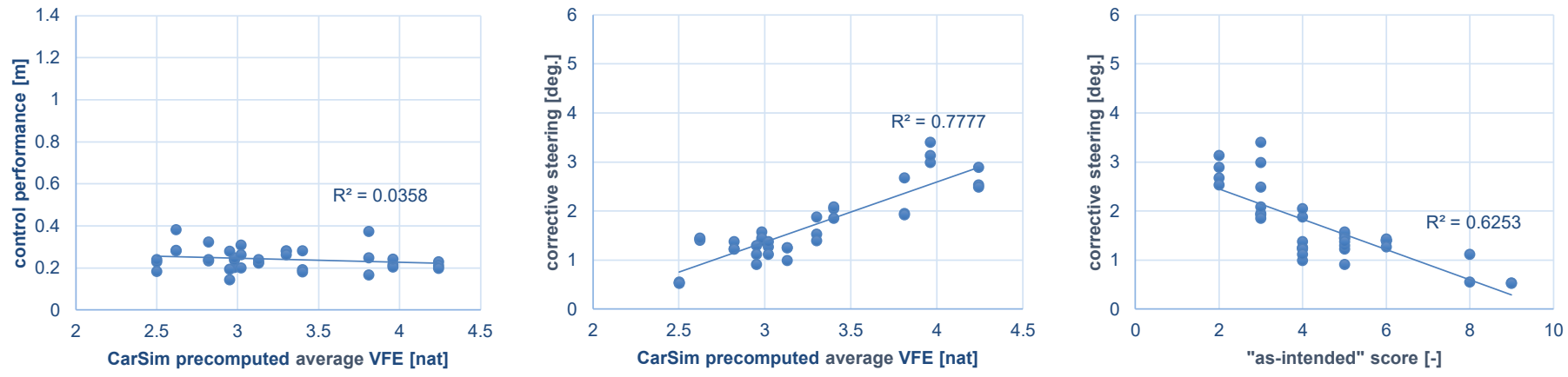

**Fig. 8.** Relationships between CarSim precomputed average VFE, control performance, corrective steering, and "as-intended" score in the expert driver.

### 3.3 Results of the Experiment by 60 General Participants

Fig. 9 (left) shows the relationship between average VFE from offline learning of each participant's driving data and their control performance, based on data from 60 general participants. A strong positive correlation was found ($r$=0.78, $p$<0.001). Fig. 9 (right) shows the relationship between the same VFE and the amount of corrective steering, with a relatively strong correlation ($r$=0.65, $p$<0.001). These results strongly support the hypothesis in Section 2.2.

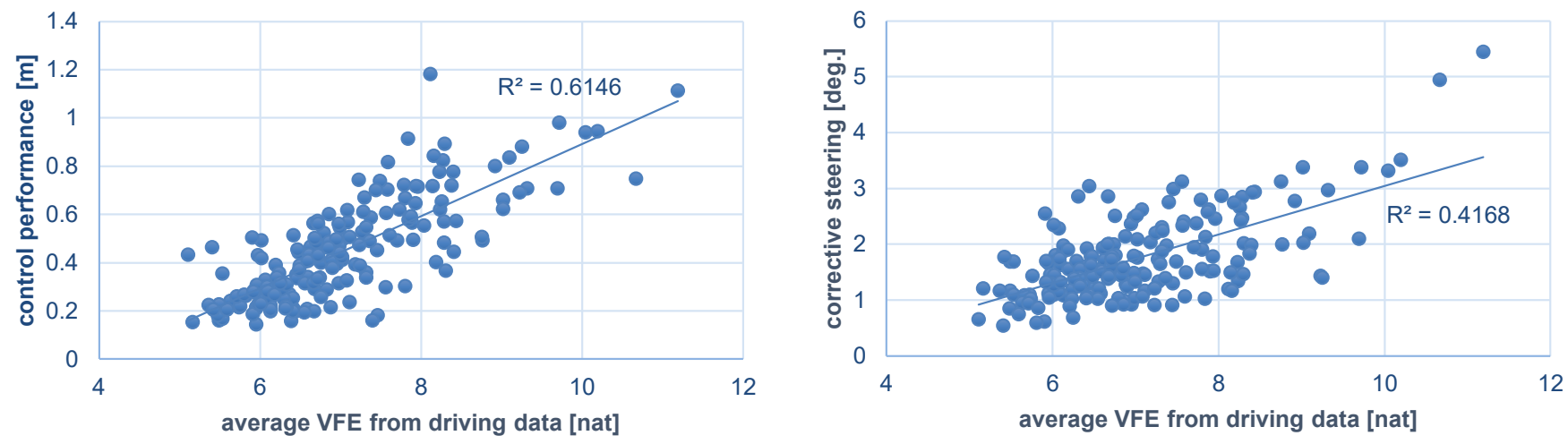

**Fig. 9.** Relationship between average VFE from offline learning of participant driving data and control performance (left) or corrective steering (right) in 60 general participants.

In contrast, Fig. 10 (left) shows the relationship between the CarSim precomputed average VFE and the "as-intended" scores given by general participants ($r$=-0.02, $p$=0.746). Fig. 10 (right) shows the relationship between the same scores and the average VFE obtained through offline learning from each participant's driving data ($r$=-0.28, $p$<0.002). Neither figure shows a clear correlation, unlike the results with the expert driver. In Fig. 10 (left), the horizontal axis represents different vehicle models, and the "as-intended" scores varied widely among general participants—even for the same vehicle. This variation was observed not only across participants, but also within the same participant across repeated trials.

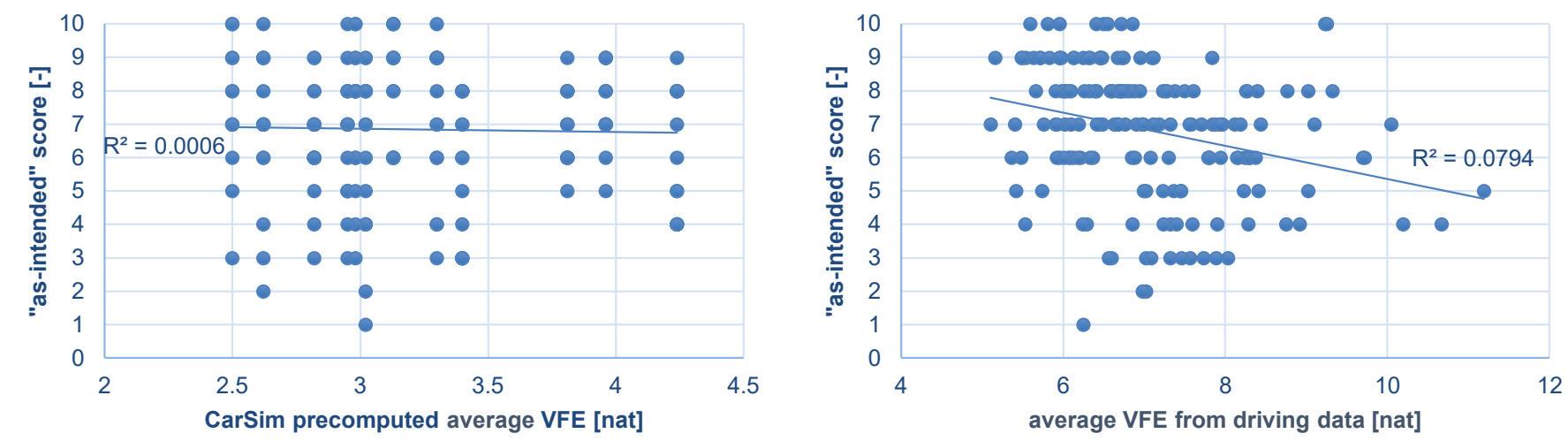

**Fig. 10.** Relationships between "as-intended" scores and average VFE in 60 general participants.

Fig. 11 further illustrates this issue. Fig. 11 (left) shows the relationship between CarSim precomputed average VFE and control performance among 60 general participants. The variation in control performance among participants is as large as or larger than that due to vehicle model differences. The resulting weak correlation (r=0.18, p<0.019) suggests that control performance is unlikely to be an indicator of the "as-intended" score. Fig. 11 (middle) shows a moderate correlation between CarSim precomputed average VFE and corrective steering (r=0.49, p<0.001). However, Fig. 11 (right) shows that corrective steering is also not a strong determinant of the "as-intended" score (r=-0.33, p<0.001). Unlike the expert driver, general participants showed no clear basis for their "as-intended" evaluations, leading to widely varying scores.

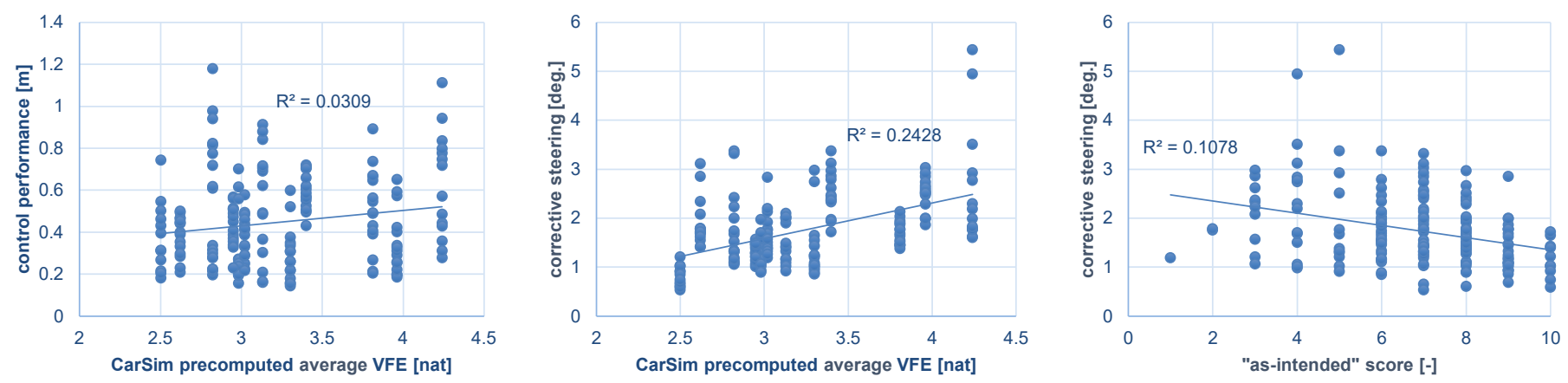

**Fig. 11.** Relationships between CarSim precomputed average VFE, control performance, corrective steering, and "as-intended" score in 60 general participants.

## 4 Discussion

Traditionally, "as-intended" controllability has been evaluated through subjective assessments by expert drivers using simulators or real vehicles. We confirmed that, while general participants showed variability in their evaluations, the expert driver's assessments consistently aligned with objective indicators such as corrective steering. This is analogous to wine tasting: trained sommeliers can evaluate reliably, whereas this is difficult for general consumers. Accordingly, we provided an explicit intention (to follow the centerline as closely as possible) and ensured objective evaluation of "as-intended" via control performance and corrective steering, even if general participants could not consciously judge it.

In this study, we introduced a novel methodology. By defining a specific task (e.g., a steering task), constructing a computational model of the brain capable of online learning within the active inference framework, and calculating VFE through offline learning using existing data, we demonstrated that the resulting VFE values correlate strongly with both reliable subjective "as-intended" scores and objective indicators (e.g., control performance or corrective steering).

These findings support the hypothesis that lower VFE reflects a higher degree of "as-intended" controllability. They also suggest that VFE values obtained through offline learning using simulation data can quantify this controllability and potentially serve as a proxy for expert drivers' subjective "as-intended" scores.

Ideally, VFE values should be derived from online learning within the active inference framework. However, we found that the learning outcomes were sensitive to factors such as the random seed used in action selection based on EFE, as well as the D vector. This sensitivity made it difficult to obtain stable and consistent VFE values for vehicle evaluation. It remains unclear whether this limitation arises from the framework itself or from the specific design of the generative model. Further studies are needed to clarify this issue.

## 5 Conclusion

Based on the methodology introduced and experimentally validated in this study, we conclude that, regardless of individual differences in driving skill or variations in vehicle dynamics, the modeling error of vehicle dynamics learned in the driver's brain—represented by VFE—plays a critical role in determining whether the driver can steer the vehicle "as-intended" from an objective standpoint.

These findings strongly support the theoretical perspective underlying active inference, which holds that the brain learns by accumulating statistical regularities in sensory inputs and acts through probabilistic inference. Furthermore, this study presents a novel approach to the long-standing challenge of quantitatively evaluating and understanding "as-intended" control in vehicle dynamics.

Importantly, the offline learning approach proposed in this study is not limited to simulation or simulator data—it can also be applied to real-world driving data. This enables consistent and objective evaluation of "as-intended" controllability regardless of the evaluator. Notably, the ability to perform such evaluations using only simulation data offers a clear advantage in the early stages of vehicle dynamics design. Future research will aim to extend this approach to real-world applications.

**Acknowledgments.** This study was initiated by Honda R&D Co., Ltd. in collaboration with The University of Tokyo. We would like to thank all participants in the driving experiments. We would also like to express our sincere appreciation to the pymdp development team for their timely completion and release of the JAX version in response to our request, which significantly contributed to the implementation of our simulation framework.

**Disclosure of Interests.** Kazuharu Kidera is employed by Honda R&D Co., Ltd. The other authors declare no competing interests.

## References

1. Tao, M., Sugimachi, T., Suda, Y., Shibata, K., Katou, D., Fukaya, T.: A study on vehicle dynamics characteristics that realize “as-intended” driving. Trans. Soc. Automot. Eng. Jpn. 48(6), 1265–1271 (2017) (in Japanese). https://doi.org/10.11351/jsaeronbun.48.1265
2. Friston, K.: The free-energy principle: a unified brain theory? Nat. Rev. Neurosci. 11(2), 127–138 (2010). https://doi.org/10.1038/nrn2787
3. Parr, T., Pezzulo, G., Friston, K.J.: Active Inference: The Free Energy Principle in Mind, Brain, and Behavior. MIT Press, Cambridge (2022). https://doi.org/10.7551/mit-press/12441.001.0001
4. Applied Intuition: CarSim. https://www.appliedintuition.com/products/carsim, last accessed 2025/05/16
5. Smith, R., Friston, K.J., Whyte, C.J.: A step-by-step tutorial on active inference and its application to empirical data. J. Math. Psychol. 107, 102632 (2022). https://doi.org/10.1016/j.jmp.2021.102632
6. Heins, C., Millidge, B., Demekas, D., Klein, B., Friston, K., Couzin, I.D., et al. : pymdp: A Python library for active inference in discrete state spaces. J. Open Source Softw. 7(73), 4098 (2022). https://doi.org/10.21105/joss.04098
7. Bradbury, J., Frostig, R., Hawkins, P., Johnson, M.J., Leary, C., Maclaurin, D., et al.: JAX: composable transformations of Python+NumPy programs. https://github.com/google/jax, last accessed 2025/05/16
8. CEDEC Digital Library: Brainwave frequency bands and information processing speed during gameplay. https://cedil.cesa.or.jp/cedil_sessions/view/2146 (in Japanese), last accessed 2025/05/16
9. Parr, T., Markovic, D., Kiebel, S.J., Friston, K.J.: Neuronal message passing using Mean-field, Bethe, and Marginal approximations. Sci. Rep. 9(1), 1889 (2019). https://doi.org/10.1038/s41598-018-38246-3